\documentstyle[12pt]{article}                                 

\begin{document}

\newcommand {\bea}{\begin{eqnarray}}
\newcommand {\eea}{\end{eqnarray}}
\newcommand {\be}{\begin{equation}}
\newcommand {\ee}{\end{equation}}

\def\IR{{\hbox{{\rm I}\kern-.2em\hbox{\rm R}}}}
\def\IH{{\hbox{{\rm I}\kern-.2em\hbox{\rm H}}}}
\def\IC{{\ \hbox{{\rm I}\kern-.6em\hbox{\bf C}}}}
\def\IZ{{\hbox{{\rm Z}\kern-.4em\hbox{\rm Z}}}}

\title{
\begin{flushright}
\begin{small}
hep-th/9805097\\
UPR/803-T \\
May 1998 \\
\end{small}
\end{flushright}
\vspace{1.cm}
Near Horizon Geometry of Rotating Black Holes
in Five Dimensions}

\author{Mirjam Cveti\v{c} and Finn Larsen\\
\small Department of Physics and Astronomy\\
\small University of Pennsylvania\\
\small Philadelphia, PA 19104 \\
\small e-mail: cvetic,larsen@cvetic.hep.upenn.edu
}
\date{ }
\maketitle

\begin{abstract}
We interpret the general rotating black holes in five dimensions 
as rotating black strings in six dimensions. In the near horizon limit
the geometry is locally $AdS_3\times S_3$, as in the nonrotating 
case. However, the global structure couples the $AdS_3$ and the $S_3$,
giving angular velocity to the $S_3$. The asymptotic geometry is exploited 
to count the microstates and recover the precise value of the 
Bekenstein-Hawking entropy, with rotation taken properly into account. 
We discuss the perturbation spectrum of the rotating black hole, and its 
relation to the underlying conformal field theory.
\end{abstract}                                

\section{Introduction}
\label{sec:intro}
In the last few years there has been substantial progress in the quantum 
description of black holes in string theory. (For reviews of this 
development see {\it e.g.}~\cite{amandareview,youmreview}.).
However, string theory has not yet offered
decisive progress on the notorious questions concerning the 
information flow in the black hole spacetime, and the physical nature of 
its singularity. The reason is that, in critical string theory, the 
internal structure of black holes is represented by world-volume 
field theories that are decoupled from gravity, and in this description
the role of spacetime geometry is unclear. However, it has recently been 
proposed that the world-volume description is in fact equivalent
to string theory in the curved background of the 
black hole~\cite{juanads,wittenads,polyakovads}. The defining examples
of this correspondance employ geometries that are products of spaces with 
constant curvature, with one factor being the Anti-deSitter spacetime (AdS). 
In these cases the world-volume theory is a conformal field 
theory (CFT), and 
the proposal can be justified in explicit computations. Even so, it is 
not yet clear whether these novel dualities will shed light on the black 
hole problems alluded to above, but this possibility is an important 
motivation for their further development.

A significant application of the correspondance between conformal
field theory and the geometry of AdS spaces is the counting of 
microstates~\cite{btzentropy,dublinbtz} of the three dimensional black 
hole of Banados, Teitelboim, and Zanelli (BTZ)~\cite{btz,btzreview}. 
The new microscopic derivation of the black hole entropy follows from 
little more than the fact that the BTZ geometry is asymptotically $AdS_3$, 
lending a surprising robustness to the result. This feature of the 
computation appears to be an important advance over the earlier work of 
Carlip~\cite{btzentrop}. Additionally, it is notable that the near horizon 
geometry of the D1-D5 bound 
state is of the form ${\rm BTZ}\times S_3$~\cite{skenderis97a}, 
allowing application of the new method to this case~\cite{btzentropy}. 
This has lead to a close connection with previous work on D-brane black 
holes~\cite{strom98a,martinecads3}. The near horizon geometry 
of three orthogonally intersecting M5-branes in $M$-theory 
is similarly ${\rm BTZ}\times S_2$, and this gives a simple relation to 
previous work on four dimensional black holes~\cite{bl98}.

The purpose of the present paper is to investigate this
construction in the context of rotating brane configurations. Our 
starting point is the most general class of black holes in five 
dimensions~\cite{cy96a}. We find the exact form of the corresponding 
six-dimensional black strings, and then take the decoupling 
limit. The resulting near horizon geometry describes near extremal 
black holes and is again of the factorized form ${\rm BTZ}\times S_3$.
Interestingly, this is explicit only in a coordinate system that is 
rotating at the same rate as the black hole. The parameters of the geometry 
gives the central charge of the effective conformal field theory, as well 
as the levels of the black hole states. A microscopic entropy can be inferred 
from these, and the result agrees precisely with the area law.
The result takes into account the precise dependence of the two 
angular momenta. Black hole entropy is discussed in sec.~\ref{sec:entropy}.

A central result of the CFT/AdS correspondence is the relation between 
the spectrum of perturbations in the $AdS_3\times S_3$ geometry, and the 
elementary excitations in the worldvolume theory of the 
$D1-D5$ system~\cite{strom98a,martinecads3,sezginads3}. 
The rotating black hole background is also locally $AdS_3\times S_3$ so
the spectrum is identical in the two cases, but in the black hole 
background it is natural to discuss the perturbations in terms of 
greybody factors. In this way the interpretation of perturbations depends 
on the boundary conditions, and thus on global issues. Specifically, when 
rotation is included, the coordinates on the sphere $S_3$ depend on the 
$AdS_3$ coordinates; and this ``twisting'' of the sphere affects the greybody 
factors in a universal manner. The perturbations of the black hole 
are discussed in sec.~\ref{sec:waveeq}, and the relation between the 
$D1-D5$ system and the black hole is the topic of the concluding 
remarks in sec.~\ref{sec:discussion}.

The BTZ background is locally $AdS_3$ and so it exhibits an 
$SL(2,\IR)_L\times SL(2,\IR)_R$ symmetry. A preferred set of coordinates can 
be defined that makes the factorized form of the symmetry manifest.
These coordinates provide a spacetime interpretation of the effective
string world-sheet. In this concrete realization the temporal and spatial 
world-sheet coordinates are naturally associated with the outer and inner 
horizons, respectively. This connection, discussed in sec.~\ref{sec:sl2},
may have universal significance.

There is a generalization of the present work to the case of
four dimensional rotating black holes and their representations as
rotating strings in five dimensions. This will be discussed in a 
separate note~\cite{cl98b}.

The paper is organized in two parts.
In  the first part, sec.~\ref{sec:entropy}, we focus on the black hole entropy:
successive subsections give the metric of the general rotating black string 
in six dimensions, find the decoupling limit and calculate the entropy.
In subsec.~\ref{sec:units} we give the translation between
macroscopic black hole parameters and the microscopic quantum numbers. 
In the second part, sec.~\ref{sec:waveeq}, we discuss black hole 
perturbations. The emphasis is on the effects of rotation, but we also 
make general remarks. We exhibit the local symmetries and recover the 
near-horizon wave equation for a minimally coupled scalar field in the 
rotating background. Finally, in sec.~\ref{sec:discussion}, we conclude with 
a discussion of the relation between the conformal field theory induced 
by the near horizon geometry of the $D1-D5$ system and that of the black 
hole.

\section{Black Hole Entropy}
\label{sec:entropy}
In this section we present the general form of the rotating black string 
in six dimensions. We then take the decoupling limit and calculate the 
microscopic entropy of rotating black holes in five dimensions,
following Strominger~\cite{btzentropy}.

\subsection{The Classical Background}
\label{sec:class}
Consider the most general rotating black holes in $N=4$ or
$N=8$ supergravity in five dimensions~\cite{cy96a}. 
The generating solution for
these configurations is given in terms its mass $M$, 2 angular momenta 
$J_{L,R}$, and 3 independent $U(1)$ charges $Q_i$. 
It is convenient to represent these physical parameters in the
parametric form:
\bea
M &=& m\sum_{i=0}^2 \cosh 2\delta_i~, \\
Q_i &=& m\sinh 2\delta_i~~~;~~i=0,1,2~,\\
J_{L,R} &=& m(l_1\mp l_2)(\prod_{i=0}^2 \cosh\delta_i \pm
\prod_{i=0}^2 \sinh\delta_i)~.
\label{eq:param}
\eea
We work in Planck units where the gravitational coupling 
constant in five dimensions
is $G_5={\pi\over 4}$. The relation to conventional string
units is given below, in sec.~\ref{sec:units}.

For the present purpose it is essential that the five dimensional black 
holes can be interpreted as rotating black strings in six dimensions. 
The six-dimensional geometry can be determined from the 
five dimensional form of the metric and the matter fields, given 
in~\cite{cy96a}. In Einstein frame the result is~\footnote{This 
corrects the result reported in~\cite{cy96c}. Also, there is a typo in 
the expression for the gauge field $A^{(1)}_\phi$ given in~\cite{cy96a}. 
The correct formula, used to derive the result given here, can be 
obtained from the $A^{(1)}_\psi$ given there, using the symmetry given 
below in eq.~\ref{eq:symm}.}:
\bea
ds^2_{6} &=& {1\over\sqrt{H_1 H_2}}
\left[ -(1-{2mf_D\over r^2}) d{\tilde t}^2 + 
d{\tilde y}^2
+H_1 H_2 f^{-1}_D {r^4 \over (r^2+l_1^2)(r^2+l_2^2)-2mr^2}dr^2 
\right.  \nonumber \\
&-&{4mf_D\over r^2} \cosh\delta_1\cosh\delta_2
(l_2\cos^2\theta d\psi + l_1\sin^2\theta d\phi)d{\tilde t} \nonumber \\
&-&{4mf_D\over r^2} \sinh\delta_1\sinh\delta_2
(l_1\cos^2\theta d\psi + l_2\sin^2\theta d\phi)d{\tilde y} \nonumber \\
&+& \left(  (1+{l^2_2\over r^2})H_1 H_2 r^2 + 
(l_1^2-l_2^2)\cos^2\theta
({2mf_D\over r^2})^2\sinh^2\delta_1\sinh^2\delta_2 \right)\cos^2\theta 
d\psi^2 \nonumber \\
&+& \left(  (1+{l^2_1\over r^2})H_1 H_2 r^2 + 
(l_2^2-l_1^2)\sin^2\theta
({2mf_D\over r^2})^2\sinh^2\delta_1\sinh^2\delta_2 \right)\sin^2\theta 
d\phi^2 \nonumber \\
 &+& \left.{2mf_D\over r^2} (l_2 \cos^2\theta d\psi +
l_1\sin^2\theta d\phi )^2
+H_1 H_2 r^2 f^{-1}_D d\theta^2
\right]~, 
\label{eq:bstring} 
\eea
where:
\bea
H_i &=& 1+ {2mf_D\sinh^2\delta_i\over r^2}~~~;i=1,2~,\\
f_D^{-1} 
&=& 1 + {l_1^2\cos^2\theta\over r^2}+{l_2^2\sin^2\theta\over r^2}~, 
\eea
and: 
\bea
d{\tilde t} &=& \cosh\delta_0 dt - \sinh\delta_0 dy~, 
\label{eq:ttilde}
\\
d{\tilde y} &=& \cosh\delta_0 dy - \sinh\delta_0 dt~.
\label{eq:ytilde}
\eea
Although the six dimensional rotating black string metric is somewhat
involved it is much simpler than the five dimensional black hole that 
it was derived from. Note that the angular parts are constrained 
by the symmetry under the simultaneous interchanges:
\be
l_1\leftrightarrow l_2~~~,~~
\phi\leftrightarrow\psi~~~,~~
\theta\leftrightarrow {\pi\over 2}-\theta~.
\label{eq:symm}
\ee
This symmetry expresses an automorphism of the algebra of rotations
$SO(4)\simeq SU(2)_R\times SU(2)_L$ that corresponds geometrically
to a reflection.

The black string metric eq.~\ref{eq:bstring} depends on the 
parameter $\delta_0$ through the boosted differentials $d{\tilde t}$
and $d{\tilde y}$ only, thus signifying a momentum along the
string. This property is by no means manifest in the five dimensional 
form of the metric given in~\cite{cy96a}; and so it serves as an 
important check on the algebra that we recover the boost invariance
in our solution. 

The two charges $Q_1$ and $Q_2$ appear symmetrically in the black string
metric, due to duality of the effective six-dimensional theory. 
They can be interpreted as the charges of a fundamental 
string (FS) wrapped around the $y$-direction and a NS5-brane wrapped 
around both the $y$-direction and the additional four 
compact dimensions. The matter fields that are needed for this 
interpretation were given in~\cite{cy96a}. Since duality transformations 
act on the matter fields, but not on the Einstein metric, we can equally 
well interpret the metric as the field created by any brane configuration 
that is dual to the pair NS5-FS. 
Specifically the $U(1)$ charges can be interpreted as D1- and D5-brane 
charges. The advantage of this interpretation is that, unlike the 
NS5-brane, the D-branes are accessible to a weakly coupled microscopic
description. For this reason the
D-brane interpretation of the solution is in fact mandatory 
for the decoupling limit taken below.

\subsection{The Near Horizon Geometry}
\label{sec:nearhor}
In some circumstances the internal structure of a black hole is  
described accurately by a field theory that couples weakly to the 
surrounding space. A precise definition of the decoupling limit 
is given by taking~\cite{juanads}:
\be
l_s\rightarrow 0~~;~
r, m, l_{1,2}\rightarrow 0~~;~
\delta_{1,2}\rightarrow\infty~,
\label{eq:decoup}
\ee
where the string length $l_s=\sqrt{\alpha^\prime}$, so that:
\be
rl_s^{-2}~~~;~
ml_s^{-4}~~~;~
l_{1,2} l_s^{-2}~~~;~
Q_{1,2} l_s^{-2}= ml_s^{-2}\sinh 2\delta_{1,2}~,
\ee
remain fixed. The decoupling limit is a near horizon approximation, because
$r\rightarrow 0$. Moreover, the ``dilute gas'' conditions
$\delta_{1,2}\rightarrow\infty$ imply that the black hole is necessarily 
near extremal~\cite{greybody}.

The metric simplifies dramatically in the limit specified
by eq.~\ref{eq:decoup}. Note, however, that the function $f_D$ does 
not simplify in this limit, and other 
features due to rotation are similarly retained. 

The near horizon geometry is:
\bea
ds^2_{6} &=&  
{r^2\over f_D\lambda^2}[-(1-{2mf_D\over r^2}) d{\tilde t}^2 + 
d{\tilde y}^2 ]
+ {\lambda^2 r^2 \over (r^2+l_1^2)(r^2+l_2^2)-2mr^2}dr^2 - 
\label{eq:nearhmetric} \\
&-&2 (l_2\cos^2\theta d\psi + l_1 \sin^2\theta d\phi)d{\tilde t}
-2 (l_1\cos^2\theta d\psi + l_2 \sin^2\theta d\phi)d{\tilde y} + 
\nonumber \\
&+& \lambda^2 (d\theta^2 + \sin^2\theta d\phi^2 + \cos^2\theta d\psi^2)~,
\nonumber
\eea
where we defined the characteristic length scale 
$\lambda\equiv (Q_1 Q_2)^{1\over 4}$.
Introducing the shift in the angular variables:
\bea
d{\tilde\psi} &=& d\psi - \lambda^{-2} (l_2 d{\tilde t}+l_1 d{\tilde y})~,
\label{eq:psitilde}
 \\
d{\tilde\phi} &=& d\phi - \lambda^{-2} (l_1 d{\tilde t}+l_2 d{\tilde y})~,
\label{eq:phitilde}
\eea
the metric becomes:
\bea
ds^2_{6} &=&  
-{(r^2+l_1^2)(r^2+l_2^2)-2mr^2\over \lambda^2 r^2 }d{\tilde t}^2 + 
{r^2\over\lambda^2} (d{\tilde y} - {l_1 l_2\over r^2} d{\tilde t} )^2 + \\
&+& 
{\lambda^2 r^2 \over (r^2+l_1^2)(r^2+l_2^2)-2mr^2}dr^2 
+\lambda^2 [d\theta^2 + \sin^2\theta d{\tilde\phi}^2 + \cos^2\theta 
d{\tilde\psi}^2 ]~.
\nonumber 
\eea
In this form it is apparent that the geometry is a direct product
of two three dimensional spaces. The angular space is a sphere $S_3$ with 
radius $\lambda$, and the geometry with cordinates $(\tilde{t},\tilde{y},r)$ 
is a BTZ black hole in an effective $2+1$ dimensional theory with 
cosmological constant $\Lambda = - \lambda^2$. 

Indeed, the metric can be written in the standard BTZ 
form~\cite{btz,btzreview}:
\bea
ds^2_6 &=& - N^2 dt^2_{\rm BTZ} + 
N^{-2}dr^2_{\rm BTZ} + r^2_{\rm BTZ} 
(d\phi_{\rm BTZ} - N_\phi dt_{\rm BTZ})^2 + 
\lambda^2 d{\tilde\Omega}^2_3 ~,\\
N^2 &=& {r^2_{\rm BTZ}\over \lambda^2} - M_3
+ {16G_3^2 J_3^2\over r^2_{\rm BTZ}}~, \\
N_\phi &=& {4G_3 J_3\over r^2_{\rm BTZ}}~,
\eea
where:
\bea
t_{\rm BTZ}&\equiv & {t\lambda\over R_y}~,\\
\phi_{\rm BTZ}&\equiv & {y\over R_{y}}~,\\
r^2_{\rm BTZ} &\equiv & 
{R_{y}^2\over \lambda^2}[r^2 + (2m-l^2_1-l^2_2)\sinh^2\delta_0
+2l_1 l_2\sinh^2\delta_0\cosh^2\delta_0]~,
\eea
and the effective 
three dimensional mass $M_3$ and angular momentum $J_3$ are:  
\bea
M_3 &=& 
{R^2_y\over \lambda^4} [ (2m-l^2_1 - l^2_2)\cosh 2\delta_0 +
2l_1 l_2\sinh 2\delta_0 ]~,\\
8G_3 J_3 &=& {R^2_y\over\lambda^3} 
[(2m-l^2_1 - l^2_2)\sinh 2\delta_0 +2l_1 l_2\cosh 2\delta_0 ]~.
\eea
We denoted the radius of the compact dimension by $R_y$.

The extremal limit of the six dimensional string is given by:
\be
m,l_{1,2}\rightarrow 0~~;~\delta_0\rightarrow\infty~,
\ee
with fixed $Q_0=m\sinh 2\delta_0$. In this limit one of the
angular momenta $J_R\rightarrow 0$, but the other $J_{L}$ 
remains finite.

It has previously been found that, in the near horizon region, 
the extremal rotating black hole preserves $1/4$ of the maximal 
supersymmetry, twice the amount that is preserved in the bulk~\cite{cfgk}. 
Now this result follows from the much simpler 
analysis of supersymmetry in the context of BTZ black holes~\cite{NSR}. 
Additionally, it follows that the near-horizon region exhibits superconformal 
invariance~\cite{cfthair2}, except for a global obstruction that can be 
removed by decompactification of the 6th dimension.

\subsection{Counting States}
\label{sec:cft}
In this section we count the microscopic states of the black hole, 
following~\cite{btzentropy}.

The effective gravitational coupling in three dimensions $G_3$ can be 
related to the gravitational coupling in five dimensions $G_5$ by 
comparing two different dimensional reductions from six dimensions, as 
in~\cite{btzentropy,bl98}. It is:
\be
{1\over G_3} = {1\over G_5}~{A_3\over 2\pi R_y}~,
\ee
where $A_3 = 2\pi^2\lambda^3$ is the area of the $S_3$. This result
is independent of the rotational parameters because the effective 
cosmological constant depends only on the charges of the branes.

The isometry group of the asymptotic $AdS_3$ induces a conformal 
field theory on the boundary at the conformal infinity of the
BTZ black hole. Its central charge is given in terms of the
cosmological constant~\cite{adsc}:
\be
c = {3\lambda\over 2G_3} = 6~{Q_1 Q_2\over R_y}~{\pi\over 4G_5}~.
\label{eq:ccharge}
\ee
Note that the central charge is also independent of angular momentum. This
suggests that the rotating black holes can be interpreted as states 
in the same conformal field theory that describes the nonrotating 
black holes.

The relation between the symmetry generators of the induced conformal 
symmetry, and the effective mass and angular momentum are:
\bea
M_3 &=& {8G_3\over\lambda}(L_0 + {\bar L}_0)~,\\
J_3 &=& L_0 - {\bar L}_0~,
\eea
where the eigenvalues of the
operators $L_0$ and ${\bar L}_0$ are the conformal dimensions
$h_L$ and $h_R$, respectively. Then Cardy's formula for the
statistical entropy~\cite{cardy}:
\be
S = 2\pi\left(\sqrt {ch_L\over 6}+\sqrt {ch_R\over 6}\right)~,
\label{eq:cardy}
\ee
gives the microscopic entropy:
\bea
S&=& {\pi\over 4G_3}
\left[
\sqrt{\lambda(\lambda M_3+8G_3 J_3)}+\sqrt{\lambda(\lambda M_3-
8G_3 J_3)}\right] \\
 &=& {\pi\over 4G_5}~\pi\sqrt{Q_1 Q_2}
\left[\sqrt{2m-(l_1-l_2)^2}~e^{\delta_0}
+\sqrt{2m-(l_1+l_2)^2}~e^{-\delta_0}\right]~.
\label{eq:microent}
\eea
The general formula for the macroscopic entropy of rotating black
holes in five dimensions is~\cite{cy96a}:
\bea
S\equiv {A_5\over 4G_5} &=&  {\pi\over 4G_5}~
2\pi m \left[ (\prod^2_{i=0}\cosh\delta_i+\prod^2_{i=0}\sinh\delta_i)
\sqrt{2m-(l_1-l_2)^2}+\right. \nonumber \\
&+&
\left. (\prod^2_{i=0}\cosh\delta_i-\prod^2_{i=0}\sinh\delta_i)
\sqrt{2m-(l_1+l_2)^2}\right]~.
\label{eq:macroent}
\eea
In the limit $\delta_{1,2}\gg 1$ this becomes:
\be
S= {\pi\over 4G_5}~\pi\sqrt{Q_1 Q_2}\left[\sqrt{2m-(l_1-l_2)^2}~
e^{\delta_0}
+\sqrt{2m-(l_1+l_2)^2}~e^{-\delta_0}\right] ~.
\label{eq:ent2}
\ee
Thus the microscopic entropy eq.~\ref{eq:microent} precisely
reproduces the macroscopic entropy eq.~\ref{eq:macroent},
in the decoupling limit eq.~\ref{eq:decoup} where the microscopic calculation 
applies. The range of parameters that are considered here is 
as general as the previous D-brane results~\cite{rotation2}.

Cardy's formula eq.~\ref{eq:cardy} can be derived from unitarity and 
modular invariance of the boundary conformal field theory. 
In the present context unitarity cannot be taken for granted;
superconformal WZW-models with the noncompact target space 
$SL(2,\IR)$ are in fact nonunitary. Thus the justification of the calculation
ultimately rests on the existence of an underlying unitary framework, 
such as the one realized in the full string theory. Despite
these caveats, we find it impressive that the method accurately 
reproduces an entropy with the complexity apparent in eq.~\ref{eq:ent2}.

\subsection{Quantization Conditions}
\label{sec:units}
It is instructive to rewrite some of the formulae in microscopic units.
Then:
\be
G_5 = {\pi\over 4}~{ (\alpha^\prime)^4 g^2\over R_1 R_2 R_3 R_4 R_y}~,
\ee
where the $R_i$ are the radii of the compact dimensions and the type IIB
string
coupling $g$ is normalized so that $g\rightarrow 1/g$ under 
S-duality.
Our convention hitherto was $G_5={\pi\over 4}$, except where $G_5$
is written explicitly. The quantization conditions 
on the D-brane charges are~\cite{dnotes}:
\bea
Q_1 &=& n_1 g ~ {\alpha^{\prime 3}\over R_1 R_2 R_3 R_4}~,\\
Q_2 &=& n_2 g \alpha^\prime~, 
\eea
where $n_{1,2}$ are the number of D1- and D5-branes, respectively.
It immediately follows from eq.~\ref{eq:ccharge} that
$c=6n_1 n_2$, as expected~\cite{strom96a}.

The quantum numbers $p$, $\epsilon$, and $j_{R,L}$ for
momentum, energy, and angular momenta, respectively, are introduced through:
\bea
Q_0 &=& m\sinh 2\delta_0 = {p\over R_y}~{4G_5\over\pi}~, \\
E  &=&  m\cosh 2\delta_0 = {\epsilon\over R_y}~{4G_5\over\pi}~,\\
J_{R,L} &=& j_{R,L}~{4G_5\over\pi}~.
\eea
Then the conformal weights can be written as: 
\be
h_{L,R} = {\lambda M_3\pm 8G_3J_3\over 16G_3}
= {\pi R_y\over 16G_5}~[ 2m -(l_1\mp l_2)^2] e^{\pm2\delta_0}
= {1\over 2}(\epsilon\pm p) - {1\over n_1 n_2} j^2_{L,R}~.
\label{eq:efflevels}
\ee
In this form it is manifest that the effective levels agree precisely 
with the results previously derived using 
$D$-branes~\cite{rotation1,rotation2}.

The quantization conditions on the angular momenta derived from the 
periodicity conditions on the angles are 
$j_{R,L}={1\over 2}(j_\phi\pm j_\psi)$, where $j_\phi$ and 
$j_\psi$ are quantized as integers~\cite{rotation1,cl97a}. 
It follows from:
\be
h_L - h_R = p -  {1\over n_1 n_2}j_\phi j_\psi~,
\ee
that the natural spacing of the conformal weights is in multiples of 
$1/n_1 n_2$. Periodicity around the $y$-direction would imply that the 
momentum quantum number $p$ is integral, but in fact $p$ is also quantized 
in multiples of $1/n_1 n_2$. Although this fractionation is suggested 
by the classical 
geometry~\cite{structure,cfthair1,gaidaads3}, it is most 
convincingly seen using D-branes~\cite{susskind96}. In this way the
presence of angular momentum makes a qualitatively important 
effect more apparent.

The {\it effective level}:
\be
N_{L,R} \equiv {c\over 6}h_{L,R} = n_1 n_2 {\epsilon\pm p\over 2} -
j^2_{L,R}~,
\ee
is designed to take into account the fractionation. The $N_{L,R}$ can be
written in the duality invariant form:
\be
N_{L,R}= 2m^3~ 
(\prod_i \cosh\delta_i\pm\prod_i \sinh\delta_i)^2~({\pi\over 4G_5})^2-
j^2_{L,R}~. 
\ee
This generalization of the effective level may account for the black hole
entropy eq.~\ref{eq:macroent} arbitrarily far from 
extremality~\cite{cy96b,fl97,cl97a}.

\section{Black Hole Perturbations}
\label{sec:waveeq}
The isometry group $SO(2,2)\simeq SL(2,\IR)_L\times SL(2,\IR)_R$ of $AdS_3$ 
can be exploited in several ways. The computation of the entropy relies 
on the fact that the BTZ black hole is {\it asymptotically} $AdS_3$, so 
that a conformal field theory is induced at the boundary
at infinity. However, 
the BTZ geometry is in fact {\it locally} $AdS_3$. This has
important consequences for the spectrum of black hole perturbations,
and for the dynamics encoded in the greybody factors. In the present 
section we discuss these issues with emphasis on the effects of rotation.
 
\subsection{The Local $AdS_3\times S_3$}
We first make the local $AdS_3$ manifest. The near horizon metric 
eq.~\ref{eq:nearhmetric} defines a quadratic form that can be diagonalized 
and written as:
\bea
ds^2_6 &=& -\lambda^{-2}~{r^2-r^2_+\over r^2_+-r^2_-}
(r_+ d{\tilde t}  - r_- d{\tilde y} )^2
+\lambda^{-2}~{r^2-r^2_-\over r^2_+-r^2_-}
(r_+ d{\tilde y}  - r_- d{\tilde t} )^2 + \nonumber \\
&+&{\lambda^2 r^2\over (r^2-r^2_+)(r^2-r^2_-)}dr^2 + \lambda^2 
d{\tilde\Omega}^2_3~,
\label{eq:diag}
\eea
where the loci $r_\pm$ of the outer and inner horizons are:
\be
r_\pm = {1\over 2}[ \sqrt{2m- (l_1-l_2)^2}\pm\sqrt{2m- (l_1+l_2)^2} ]~.
\ee
The eigenvectors of the metric are parametrized by the dimensionless 
differentials:
\be
d\tau = \lambda^{-2} (r_+ d{\tilde t} - r_- d{\tilde y})
~~~;~~
d\sigma= \lambda^{-2}(r_+ d{\tilde y} - r_- d{\tilde t})~,
\label{eq:tausigmadef}
\ee
and the dimensionless radial variable $\rho$ is introduced through:
\be
{1\over 2}\cosh 2\rho \equiv
{r^2 - {1\over 2}(r^2_+ + r^2_-)\over r^2_+ - r^2_-}
\equiv x~~~;r\geq r_{+}~,
\label{eq:xdef}
\ee
with the coordinate $x$ defined for later use. Then the metric becomes:
\be
ds^2_6 = \lambda^2 [-\sinh^2\rho ~d\tau^2 + \cosh^2\rho ~d\sigma^2
+ d\rho^2 + d{\tilde\Omega}^2_3 ]~.
\label{eq:xmet}
\ee
In this form it is manifest that the geometry is locally $AdS_3\times S_3$
outside the horizon. The result also holds inside the horizon, as can 
be shown using alternative definitions of the radial coordinate $\rho$
that apply in different patches. Since the local geometry is just
$AdS_3\times S_3$, the origin of nontrivial causal structure is of 
purely global nature~\cite{geom21}. The rotation similarly does
not affect the local structure, but the shifts of the angular coordinates 
eqs.~\ref{eq:psitilde}--\ref{eq:phitilde} tie the $S_3$ 
to the $AdS_3$. We explore their precise effect in more detail
in the following.

\subsection{Black Hole Perturbations}

An important consequence of the local $AdS_3\times S_3$ form of the
metric is that the spectrum of black hole perturbations is organized 
into multiplets of the superconformal algebra. This allows a complete 
classification of all perturbations, as carried out in~\cite{sezginads3}. 
The spectrum of perturbations follows from local properties of $AdS_3$,
and so it is identical for the entire class of black holes considered here.

In the present context the perturbations are naturally interpreted
as test fields that interact with the black hole background. The wave 
function of the perturbations then gives the greybody factor, 
expressing the form factor of the Hawking radiation as function of 
particle quantum numbers, such as energy and spin, and
of the black hole 
parameters~\cite{greybody,cgkt,greybody2,gubser,hosomichi,cl97d}\footnote{In 
the case of minimally coupled scalars the wave function on $AdS_3\times S_3$
can be matched directly on to the asymptotic Minkowski space,
but in general the geometry interpolating between the near 
horizon region and the asymptotic Minkowski space results in further 
distortions~\cite{krasnitz97}. See~\cite{krasnitz98,mathur98} for very 
recent discussions in a related context.}. The greybody factors provide 
a semiclassical testing ground for dynamical properties. In the special 
case of minimally coupled scalars in the S-wave they agree precisely with 
calculations in string theory~\cite{mathur}. The microscopic processes
responsible for other modes can be modelled in terms of an 
effective string theory with dynamics that reproduces the black hole 
greybody factors qualitatively~\cite{cgkt,greybody2}. 

It is the
{\it local} $AdS_3\times S_3$ geometry of the near horizon that makes 
the effective string description possible. Moreover, this connection has 
made it feasible to complete the list of conformal dimensions that was 
previously known only in part~\cite{sezginads3}.

The most general black hole depends on $6$ parameters (given in 
eq.~\ref{eq:param}), but only $4$ parameters remain in the decoupling 
limit specified in eq.~\ref{eq:decoup}. They can be chosen as
$ml_s^{-4},\delta_0, l_{1,2}l_s^{-2}$, or $M,Q_0,J_{L,R}$, but for 
the present purpose it is more convenient to choose them as the 
potentials:
\bea
\beta^{L,R} &=& {2\pi\lambda^2 e^{\mp\delta_0}\over
\sqrt{2m-(l_1\mp l_2)^2}} = 
{2\pi\sqrt{n_1 n_2}\over 
\sqrt{{1\over 2}(\epsilon\pm p) n_1 n_2 - j^2_{L,R}}}~{R_y\over 2}~,
\label{eq:beta}
\\
\beta_H\Omega^{L,R} &=&{2\pi (l_1\mp l_2)\over\sqrt{2m-(l_1\mp l_2)^2}}
= {2\pi j_{L,R}\over\sqrt{{1\over 2}(\epsilon\pm p) n_1 n_2 - j^2_{L,R}}}
~.
\label{eq:omega}
\eea
The $\beta^{L,R}$ are conjugate to the left and right moving energy 
along the string, respectively; and the $\beta_H\Omega^{L,R}$ are conjugate to
the two independent angular momenta $J_{L,R}$. 

The rotating background breaks the rotational invariance so the
wave function depends nontrivially on the azimuthal quantum
numbers $m_{L,R}$. However, the rotational invariance is restored 
in the shifted coordinates eqs.~\ref{eq:psitilde}-\ref{eq:phitilde}:
\be
{1\over 2}({\tilde\psi}\pm{\tilde\phi})
={1\over 2}(\psi\pm\phi)- {l_1\pm l_2\over\lambda^2}~e^{\mp\delta_i}
(t\pm y).
\label{eq:shifts}
\ee
Simultaneous translations of the $\psi\pm \phi$ and $t\pm y$ that 
can be absorbed in translations of ${\tilde\psi}\pm{\tilde\phi}$ 
must leave the wave function invariant, except for an overall phase. 
However, the system may transform nontrivially under translations
that leave ${1\over 2}({\tilde\psi}\pm{\tilde\phi})$ fixed. 
The wave function is written in general as:
\bea
\Phi & \equiv &\Phi_0(r)~\chi(\theta)~
e^{-i\omega t+iq y + im_\phi\phi+im_\psi\psi} \nonumber \\
&=& \Phi_0(r)~\chi(\theta)~e^{-i\omega_R (t+y)-i\omega_L (t-y)
+im_R(\phi+\psi)+im_L(\phi-\psi)}~,
\label{eq:ansatz}
\eea
where $m_{R,L}\equiv {1\over 2}(m_\phi\pm m_\psi)$ and 
$\omega_{R,L}\equiv {1\over 2} (\omega\mp q)$. Note that the coefficient of 
the second term in eq.~\ref{eq:shifts} is the ratio of 
eqs.~\ref{eq:omega} and~\ref{eq:beta}. Translations in $\psi\pm \phi$ 
and $t\pm y$ that leave  ${1\over 2}({\tilde\psi}\pm{\tilde\phi})$ 
invariant are therefore conjugate to
$\beta^{L,R}\omega_{L,R}- m_{L,R}\beta_H\Omega^{L,R}$.
Thus the entire dependence of the radial wave function $\Phi_0$
on $m_{L,R}$ and 
$\beta_H\Omega^{L,R}$ can be taken into account by the
shifts:
\be
\beta^{L,R}\omega_{L,R}\rightarrow 
\beta^{L,R}\omega_{L,R}- m_{L,R}\beta_H\Omega^{L,R}~.
\label{eq:blrshifts}
\ee
This rule gives the exact wave functions in the rotating background,
when the nonrotating ones are known. It is valid for all fields in the 
near horizon region, without regard to the details of their couplings. 

\subsection{The Scalar Wave Equation}
Let us make this discussion explicit in the case of a minimally 
coupled scalar field. We use the original coordinates $t,y$ and
the radial variable $x$, introduced in eq.~\ref{eq:xdef}. Then 
metric is:
\bea
ds^2_6 &=& \lambda^{-2} 
\left[- (x-{1\over 2})~ (r_+ d{\tilde t} - r_- d{\tilde y})^2
+ (x+{1\over 2})~ (r_+ d{\tilde y} - r_- d{\tilde t})^2 \right] + 
\label{eq:xwaveeq} \\
&+& \lambda^2 \left[{1\over 4x^2 - 1} dx^2 + 
d\theta^2 + \cos^2 \theta 
d{\tilde\psi}^2 + \sin^2 \theta d{\tilde\phi}^2 \right]~,
\nonumber
\eea
where $({\tilde t}, {\tilde y})$ 
and  $({\tilde\psi}, {\tilde\phi})$
are defined in eqs.~\ref{eq:ttilde}--\ref{eq:ytilde} and 
eqs.~\ref{eq:psitilde}--\ref{eq:phitilde}, respectively.
Inserting the {\it ansatz} for the wave function (eq.~\ref{eq:ansatz}) 
into the Klein-Gordon equation:
\be
{1\over\sqrt{-g}}\partial_\mu (\sqrt{-g}g^{\mu\nu}\partial_\nu \Phi) = 
\mu^2 ~,
\label{eq:kg}
\ee
we find:
\bea
&~&\left[{\partial\over\partial x}(4x^2-1){\partial\over\partial x}
+{1\over x-{1\over 2}}
({\beta^R\omega_R+\beta^L\omega_L-m_R\beta_H\Omega^R
-m_L\beta_H\Omega^L \over 2\pi} )^2 - \right.  \nonumber \\
&-&\left. {1\over x+{1\over 2}}
({\beta^R\omega_R-\beta^L\omega_L-m_R\beta_H\Omega^R
+m_L\beta_H\Omega^L \over 2\pi} )^2 \right]\Phi_0 = 
(\Lambda+\lambda^2\mu^2)\Phi_0~,
\label{eq:geneq}
\eea
after somewhat lengthy transformations. The eigenvalues of the angular 
Laplacian:
\be
{\hat\Lambda} =
- {1\over \sin 2\theta}{\partial\over\partial\theta}
\sin 2\theta{\partial\over\partial\theta}
-{1\over \sin^2 \theta}{\partial^2\over\partial\phi^2}
-{1\over \cos^2 \theta}{\partial^2\over\partial\psi^2}~,
\label{eq:flatlap}
\ee
were denoted $\Lambda$ and take the form $\Lambda=l(l+2)$ 
where $l=0,1,\cdots$. The wave equation eq.~\ref{eq:geneq} agrees 
with the near horizon limit of the general one given in~\cite{cl97a},
except that there the {\it ansatz} for the wave function did not allow 
dependence on the compact coordinate $y$~\footnote{An equation that
applies in the general nonextremal case and includes the 
dependence on the compact coordinate $y$ can be obtained from the master 
equation given in~\cite{cl97a} without extensive calculations, 
by exploiting boost invariance in the $y$ dimension.}. The present 
generalization gives an even more symmetric result.

It is immidiately apparent from the form of the wave equation that its 
solutions depend on $m_{L,R}$ and $\beta_H\Omega^{L,R}$ only through the
prescriptions eq.~\ref{eq:blrshifts}. The solution to the
near horizon wave equation is a hypergeometric function found
in~\cite{greybody2,gpartial,mpartial} (present notation is used 
in~\cite{cl97a}).

The recent work on black hole greybody factors has focussed on 
massless fields, but here a mass $\mu$ is included in the Klein-Gordon
equation, eq.~\ref{eq:kg}.
This leads to a constant on the right hand side of the radial 
equation eq.~\ref{eq:geneq} and so the effect of the mass can be absorbed 
in the conformal dimension. The reason that the mass term does not change 
the form of the radial equation is that the determinant of the metric 
is independent of the radial variable. This property is nontrivial: 
for example, massive fields that couple to the Einstein metric in five 
dimensions experience a potential induced by the effective dilaton.
These fields are therefore more complicated than their six dimensional 
analogues considered here. It is possible that this observation points 
towards a simple spacetime description of string states that are massive 
from the six-dimensional point of view. 

In rotating backgrounds there are not in general enough isometries 
to guarantee that the variables can be separated~\footnote{We would like 
to thank G. Gibbons for reminding us of this fact.}. It is therefore 
a surprise that this is possible for massless minimally coupled 
scalars in the five dimensional black hole geometry~\cite{cl97a}. 
The statements made in the preceding paragraph can be verified
even for the general nonextremal metric eq.~\ref{eq:bstring}. Thus
we find that separation of variables remains true 
for {\it massive} scalars, when the mass is measured by the Einstein 
metric in six dimensions. These properties are precisely analogous to 
those of Kerr black holes in four dimensions. This is encouraging for 
the hope that the rotating black holes in string theory admit conserved 
Killing-Stackel tensors and Killing-Yano spinors that are analogous to 
those of the Kerr black hole 
(for some discussion and references see~\cite{susysky,cl97c}). 

\subsection{The local $SL(2,\IR)_L\times SL(2,\IR)_R$}
\label{sec:sl2}
The explicit form of the local $SL(2,\IR)_L\times SL(2,\IR)_R$ generators
were inferred from the wave equation in~\cite{cl97a}, without realizing 
the connection to the BTZ black hole. The result agrees with the
one found directly in the BTZ geometry~\cite{geom21}, except for the 
modification due to rotation. It can be written\footnote{We change the 
signature from the Euclidian one used in~\cite{cl97a} by taking 
$R^{\rm here}_\pm (\tau,\sigma)=
iR^{\rm there}_\mp (i\tau,i\sigma)$ and 
$R^{\rm here}_3(\tau,\sigma)=
iR^{\rm there}_3 (i\tau,i\sigma)$.}:
\bea
R_\pm &\equiv & R_1\pm iR_2
={1\over 2}e^{\pm (\tau+\sigma)} [\mp \partial_\rho +
(\coth\rho~\partial_\tau +\tanh\rho~\partial_\sigma)]~, \nonumber \\
R_3 &=& {1\over 2}({\partial\over\partial\tau}+{\partial\over\partial\sigma})
\label{eq:sl2}~,
\label{eq:gen}
\eea
and the $SL(2,\IR)$ algebra is normalized so:
\be
[R_3 , R_\pm ] =  \pm R_\pm ~~~;~~ [R_+ , R_- ] = -2R_3~. 
\ee 
The generators $\vec{L}$ of the $SL(2,\IR)_L$ are found
by taking $\sigma\rightarrow -\sigma$. They commute with the $\vec{R}$.
The effect of rotation is taken into account by evaluating the
derivatives {\it at fixed value of} $(\tilde{\psi},\tilde{\phi})$.

The $(\tau,\sigma)$ are the coordinates that are acted on in 
a simple way by the local $SL(2,\IR)_L\times SL(2,\IR)_R$. It is 
therefore natural to interpret these variables as the spacetime
realizations of the effective string world-sheet coordinates. 
The $\tau$ and the $\sigma$ are proportional to the $t$ and $y$,
respectively, for a static string wrapped around the $y$ direction. 
When the string has a momentum along $y$ the world-sheet coordinates 
the $(\tau, \sigma)$ are proportional to the boosted 
coordinates ${\tilde t}$ and ${\tilde y}$. However, in the case 
of a rotating string there is no obvious geometrical interpretation 
of the linear relation eq.~\ref{eq:tausigmadef}
between $(\tau,\sigma)$ and $(t,y)$.

There is a striking relation between $(\tau, \sigma)$ and the black hole 
horizons. In the metric eq.~\ref{eq:xwaveeq} we see that $g_{\tau\tau}$ 
vanishes precisely once, at the outer horizon, and $g_{\sigma\sigma}$ 
similarly vanishes once, at the inner horizon. 
Thus, upon embedding into one dimension higher, the two well known 
zeros of the $g_{tt}$, at the two horizons, have been split symmetrically 
between the remaining time-like, and the additional space-like 
coordinate. This structure is reflected in the wave equation 
eq.~\ref{eq:geneq} by the presence of simple poles at both 
horizons, with an apparent symmetry between their coefficients. The
wave equation 
implies that the potentials conjugate to L- and R- moving energy
$\beta_{L,R}$, given in eq.~\ref{eq:beta}, are related to
the surface accelerations $\kappa_{\pm}$ at the outer and inner
horizons, respectively, as~\cite{cl97a}:
\be
\beta_{L,R} = {2\pi\over\kappa_{+}}\mp {2\pi\over\kappa_{-}}~.
\label{eq:betarldef}
\ee
It is reasonable that the effective string theory of black holes
treats the effective world-sheet coordinates 
$(\tau,\sigma)$ symmetrically, but it is not obvious that 
this should imply a symmetry between the two horizons of the black hole. 
We suspect that this structure may have important implications.

The coordinates $(\tau,\sigma)$ exhibit nontrivial
global properties, due to the periodicity of 
$y~$\footnote{Note, however, that this subtlety
does not affect the differentials $(d\tau,d\sigma)$ and the differential
operators used in the generators eq.~\ref{eq:sl2}.}. It is natural to 
take both $t$ and $y$ compact. The generators eq.~\ref{eq:gen} determine 
the periodicity of $\sigma$ as $2\pi i$, and then the relation:
\be
\tau\pm\sigma = \lambda^{-2} (r_+ \mp r_-)e^{\mp\delta_0}(t\pm y)
= {2\pi\over\beta_{L,R}}~(t\pm y)~,
\ee
gives the complex structure of Euclidianized spacetime 
$z=(y+it)/2\pi R_y$ in terms of the black hole parameters as:
\be
\tau_{\rm modulus} = ({2\pi\over\kappa_+} + i{2\pi\over\kappa_-})
{i\over 2\pi R_y}~. 
\ee
In this way the two parameters that specify the global structure in spacetime 
are mapped to a complex modulus of the effective string. The remaining 
two black hole parameters, associated with rotation, are similarly 
associated with the inequivalent embeddings of the two $SU(2)$ 
algebras into the isometry group of the sphere $S_3$. Thus the global 
structure is parametrized by a total of four moduli, both in spacetime 
and on the effective string world-sheet.

\section{Discussion}
\label{sec:discussion}
In string theory black holes are described as quantum states 
that are constructed by exciting a fundamental ground 
state~\cite{strom96a}. A given black hole is described by a projection 
on the highly excited states that are consistent with the specified 
left moving (L) and right moving (R) energy and the (L and R) angular 
momenta. The projected theory forms a legitimate CFT but it is only its 
finite excitations that can be considered, or else the state belongs to 
a sector that is better described in terms of different projection. Thus 
the four parameters that describe the projections act as moduli.

On the spacetime side of the AdS/CFT correspondance 
the fundamental ground state is identified with the $AdS_3$ geometry 
of the $M=-1$ BTZ black hole~\cite{NSR,btzentropy}, and the excitations 
are general multiparticle states in the Fock space, constructed over the 
perturbative spectrum of the 
$AdS_3\times S_3$~\cite{juanads,strom98a,martinecads3}. The present paper,
and the works on greybody factors, considered the actual black hole 
spacetime, rather than the vacuum geometry.
The geometry remains locally $AdS_3\times S_3$ in the black hole spacetime, 
just as in the vacuum, so the induced CFTs have the same spectrum in 
the two cases. However, they are not equivalent: the boundary conditions 
satisfied by the wave functions that describe the perturbations are in 
one--to--one correspondance with the parameters that describe the black 
hole, and thus with the moduli of the effective string.
The ``fundamental'' CFT is just a specific point in moduli space, 
albeit one that defines the vanishing of entropy.

It is inherent to this discussion that the black hole microstates 
are ``made out of'' excitations of a different background that is not 
itself a black hole. Thus it is not meaningful to ask where in
the black hole spacetime the microstates reside, because the geometry 
is itself an emerging phenomenon that is only present in the projected
theory with finite moduli turned on. This picture therefore resolves the 
obvious tension between the no hair theorem, and the counting of black hole 
microstates.

An important aspect of the black hole puzzles is that general relativity 
applies whenever the curvature is small, and specifically in the horizon
region of large black holes. This condition can only be met in the
backgrounds described by highly excited states, and hence general relativity 
should always be compared with the effective string CFT, rather than the 
fundamental one. Moreover, the effective string CFT reduces to general 
relativity when curvatures are small, so all is in order.

However, in this framework, information flow must be analyzed in terms
of the quasistatic evolution on the moduli space, and the resolution
of the black hole singularities requires a geometric 
interpretation of the CFT that persists in the strong coupling region. 
It is not presently clear what techniques would allow such studies so
we are not yet in a position to answer the central puzzles posed by the 
black holes.

\vspace{0.2in} {\bf Acknowledgments:} 
We would like to thank V. Balasubramanian, R. Leigh, J. Lykken 
and J. Maldacena for discussions. This work is supported in part 
by DOE grant DOE-FG02-95ER40893.

\end{document}